\documentclass[]{aa}

\usepackage{graphicx}
\usepackage{adjustbox}
\usepackage{txfonts}
\usepackage{url}
\usepackage{relsize}
\usepackage{color}
\usepackage{amstext}

\begin{document}

\def\simlt{\mathrel{\rlap{\lower 3pt\hbox{$\sim$}}\raise 2.0pt\hbox{$<$}}}
\def\simgt{\mathrel{\rlap{\lower 3pt\hbox{$\sim$}} \raise 2.0pt\hbox{$>$}}}

\def\lsim{\,\lower2truept\hbox{${< \atop\hbox{\raise4truept\hbox{$\sim$}}}$}\,}
\def\gsim{\,\lower2truept\hbox{${> \atop\hbox{\raise4truept\hbox{$\sim$}}}$}\,}

\def\beq{\begin{equation}}
\def\eeq{\end{equation}}

   \title{Predictions for the diffuse cosmic dipole \\ at radio frequencies from reionization imprints}

    \author{T. Trombetti\inst{1,2}\fnmsep\thanks{e-mail:trombetti@ira.inaf.it}
                     \and
           C. Burigana\inst{1,3,4}
           }

    \institute{INAF, Istituto di Radioastronomia, Via Piero Gobetti 101, I-40129 Bologna, Italy
    \and
    CNR, Istituto di Scienze Marine, Via Piero Gobetti 101, I-40129 Bologna, Italy
    \and
    Dipartimento di Fisica e Scienze della Terra, Universit\`a di Ferrara, Via Giuseppe Saragat 1, I-44122 Ferrara, Italy
    \and
    INFN, Sezione di Bologna, Via Irnerio 46, I-40127 Bologna, Italy}

   \date{Received ...; accepted ...}

 \abstract{The cosmological reionization and thermal history, following the recombination epoch and the dark age, can be studied at radio frequencies through the tomographic view offered by the redshifted 21cm line and the 
integrated information offered by the diffuse free-free emission, coupled to the Comptonization distortion, which is relevant at higher frequencies.
For these types of signals, current theoretical predictions span a wide range of possibilities. The recent EDGES observations of the monopole disagree with the typical standard models and call, if confirmed, for non-standard physical processes and/or for an early population of extragalactic sources producing a remarkable radio background at high redshifts that is almost consistent with the ARCADE 2 claim of a significant excess of cosmic microwave background (CMB) absolute temperature at low frequency.
These signatures can be observed both in global (or monopole) signal and fluctuations from very large to small angular scales. 
The peculiar motion of an observer with respect to an ideal reference frame, at rest with respect to the CMB, produces boosting effects in several observable quantities. They are remarkable in the
anisotropy patterns at low multipoles, particularly in the dipole, with frequency spectral behaviours depending on the spectrum of the monopole emission, as previously studied in the context of CMB spectral distortions.
We present here a novel investigation of this effect at radio frequencies, aimed at predicting the imprints expected in the redshifted 21cm line signal and in the diffuse free-free emission plus the Comptonization distortion for several representative models. Furthermore, we consider the same type of signal, but as expected from the cosmological (CMB plus potential astrophysical signals) radio background determining the offset for 21cm redshifted line.
The combination of the four types of signal and their different relevance in the various frequency ranges is studied. This approach of linking monopole and anisotropy analyses,
can be applied on all-sky or relatively wide sky coverage surveys as well as to a suitable set of sky patches. By relying only on the quality of interfrequency and relative data calibration, the approach
in principle by-passes the need for precise absolute calibration, which is a critical point of current and future radio interferometric facilities.}

   \keywords{Diffuse radiation; cosmic background radiation; dark ages, reionization, first stars.}

   \maketitle


\section{Introduction}

The recent results from the {\it Planck} satellite support cosmological reionization scenarios that are almost compatible with astrophysical models for the evolution of structure, galaxy, and star formation. 
On the other hand, the full understanding of the reionization and thermal history of the Universe since the recombination epoch is still far from consolidated, and it is important to distinguish among the various models that are compatible with current data.  
The radio- to sub-millimetre background provides a very important window for studying this process in a global approach and achieving a complete comprehension of the involved photon and energy sources. 
We mainly focus here on the information encoded at radio frequencies, where the redshifted 21cm line from neutral hydrogen offers a tomographic view of the process, and the 
diffuse free-free (FF) emission provides a fundamental information integrated over the relevant epochs. Furthermore, the precise understanding of the radio extragalactic background is linked to the comprehension of these signals.

The peculiar motion of an observer with respect to an ideal reference frame, set at rest with respect to the cosmic microwave background (CMB),
produces boosting effects in the anisotropy patterns at low multipoles, in particular in the dipole, with frequency spectral behaviour related to the spectrum of the corresponding monopole emission.
The possibility of studying the dipole anisotropy spectrum to extract information on the CMB spectral distortions has originally been proposed \citep{1981A&A....94L..33D} with the aim of constraining energy dissipations that may have occurred in the cosmic plasma at different epochs, focussing on Bose-Einstein-like and Comptonization distortions. This topic has recently been investigated
in the context of future CMB anisotropy missions \citep{2015ApJ...810..131B}, including the possibility of better constraining the cosmic infrared background (CIB) spectrum \citep{2016JCAP...03..047D}, and
considering both these topics including instrumental performance as well as limitations by foregrounds and relative calibration uncertainties, also through numerical simulations \citep{2018JCAP...04..021B}.
The possibility of applying this differential approach to the analysis of the redshifted 21cm line has been advanced by \cite{2017PhRvL.118o1301S}, while 
\cite{2018ApJ...866L...7D} proposed to exploit the corresponding diurnal pattern in drift-scan observations.

In this work we present the predictions for the diffuse cosmic dipole, mainly at radio frequencies, but in some cases also in the microwave and sub-millimetre domain, 
from four types of imprints and their combinations expected from (or associated with) cosmological reionization: the diffuse FF emission and the coupled Comptonization distortion, the redshifted 21cm line, and the 
radio extragalactic background for a set of models that spans currently plausible possibility ranges. 

In Sect. \ref{sec:monmod} we present and discuss the models we adopted for the monopole spectra, including different extragalactic residual radio backgrounds from 
source contribution subtractions at different detection thresholds. In Sect. \ref{sec:diff_approach} we briefly describe the adopted basic formalism of the differential approach. 
Our main results for each specific type of signal are presented in Sect. \ref{sec:each}, and those regarding the dipole spectra expected from combinations of these imprints are described in Sect. \ref{sec:combi}.
Finally, in Sect. \ref{sec:conclu} we draw our main conclusions.

\section{Monopole models}
\label{sec:monmod}

\subsection{Joint free-free diffuse emission and Comptonization distortion}
\label{sec:FF}

The cosmological reionization process associated with the primeval formation phases of bound structures, which are sources of photon and energy production, generates coupled Comptonization distortions 
because of electron heating \citep{1972JETP...35..643Z} and FF distortions. The key parameters for characterizing these effects as functions of time are the Comptonization parameter,
\begin{equation}\label{eq:uComp}
u(t)=\int_{t_{\rm i}}^{t} [(\phi-\phi_{\rm i})/\phi] (kT_{\rm e}/m c^2) n_{\rm e} \sigma_{\rm T} c dt \, 
,\end{equation}
\noindent and the FF distortion parameter $y_{B} (t,x),$
\begin{eqnarray}
y_{B}(t,x) & = & \int_{t_{h}}^{t} (\phi - \phi_{i}) \phi^{-3/2} g_{B} (x, \phi) K_{0B} dt  \nonumber
\\ & = & \int_{z}^{z_{h}} (\phi - \phi_{i}) \phi^{-3/2} g_{B} (x, \phi) K_{0B} t_{exp} \frac{dz}{1+z} \, ,
\label{eq:free}
\end{eqnarray} 
\noindent 
where $z$ is the redshift, $T_e$ the electron temperature, $t_{exp}$ the cosmic expansion time, and
$K_{0B} (z) = K_B (z) / \phi^{-7/2}$, where $K_B (z)$ is expressed by
\begin{align}
K_B (z) & = {8\pi \over 3} { { e^6 h^2 n_e (n_H+4n_{He}) } \over { m (6\pi m k T_e)^{1/2} (k T_e)^3 } } \nonumber \\ &
\sim 2.6 \cdot 10^{-25} \phi^{-7/2} (T_0 / 2.7 {\rm K})^{-7/2} (1+z)^{5/2} {\hat{\Omega}_{b}}^2 \, \chi_e^2 \, {\rm sec}^{-1} \, .
\label{eq:KB}
\end{align} 
In the above equations $x=h\nu/(kT_r)$ and $T_r=T_0(1+z)$ are the redshift invariant dimensionless frequency and the redshift-dependent effective temperature of the CMB,
$g_{B} (x, \phi)$ is the Gaunt factor, $\phi = T_e/T_r$ and $\hat{\Omega}_{b} = \Omega_{b} h_{50}^2$, with 
$h_{50} = {H_{0}}/({50\, \rm{km\, sec^{-1} Mpc^{-1}})} $, $H_0$ is the Hubble constant, and ${\Omega}_{b}$ 
is the baryon density parameter. Here, $T_0 = (2.72548 \pm 0.00057)$ K \citep{2009ApJ...707..916F} is the current effective temperature of the CMB in the blackbody spectrum approximation, that is, such that $aT_0^4$ gives the current CMB energy density. In Eqs. \eqref{eq:uComp} and \eqref{eq:KB}, $n_e$ is the density of free electrons. A mixture of hydrogen and helium 
with densities in the ionized state $n_H$ and $n_{He}$ is assumed for simplicity in the first equality of Eq. \eqref{eq:KB}, while the approximation in
the second equality of Eq. \eqref{eq:KB} denotes the dependence on the ionization degree, $\chi_e$. A more accurate formula according to the convention adopted to define $\chi_e$ should include 
the atom fractions in different ionization states (see e.g. \cite{2014MNRAS.437.2507T}), but this is not relevant in the context of this paper. 
In the above equations, $k$ and $h$ are the Boltzmann and Planck constant, $m$ is the electron mass, $c$ the speed of light, $\sigma_{\rm T}$ the Thomson cross section, and
$\phi_{\rm i} = \phi(z_{\rm i}) = (1+ \Delta \epsilon/ \varepsilon_{\rm i})^{-1/4} \simeq 1-u$ is
the ratio between the equilibrium matter temperature and the radiation temperature at the beginning of the heating process (i.e. at $z_{\rm i}$).

The distorted photon occupation number is well approximated by
\begin{equation}\label{eq:etaC}
\eta^{\rm FF+C} \simeq \eta_{\rm i} + u {x / \phi_{\rm i} {\rm exp}(x/\phi_{\rm i}) \over [{\rm exp}(x/\phi_{\rm i}) - 1]^{2}}  \left ( {x/\phi_{\rm i} \over { {\rm tanh}(x/2\phi_{\rm i}) }} - 4 \right) + \frac{y_{B}(x)}{x^{3}}
,\end{equation}
\noindent where $\eta_{\rm i}$ is the initial photon occupation number (i.e. before the beginning of the dissipation process). In the following, $\eta_i$ is assumed to have a Planckian distribution. At low frequencies $\eta^{\rm FF+C}$ can be approximated by the expression:
\begin{equation}\label{eq:etaCapp} 
\eta^{\rm FF+C} \simeq \eta_{\rm i} + \frac{y_{B}(x)}{x^{3}} - \frac{2u}{x/\phi_{\rm i}} \simeq \frac{1-3u}{x} + \frac{y_{B}(x)}{x^{3}} \, .
\end{equation}

Eqs. \eqref{eq:uComp},  \eqref{eq:free}, and \eqref{eq:etaC} hold at any redshift, provided that the two distortion parameters $y_{B}$ and $u$ are integrated over the corresponding redshift interval. 
Performing the integral over the relevant cosmic times and in the limit of small energy injections (according to current limits), we obtain the relation between $u$ and the global fractional energy exchange 
between matter and radiation in the cosmic plasma: $u \simeq (1/4) \Delta\varepsilon/\varepsilon_{\rm i}$.

Defining the thermodynamic temperature, $T_{\rm th}(\nu)$, as the temperature of the blackbody having the same $\eta(\nu)$ at the frequency $\nu$,
\begin{equation}
      T_{\rm th} (\nu) ={h\nu\over k} {1 \over \ln(1+1/\eta(\nu))} \, ,
    \label{eq:t_therm}
\end{equation}
\noindent
the CMB spectrum can be approximated at low frequencies by the relation
\begin{equation}
T_{\rm th}^{\rm FF+C}(x) \simeq \left ( \frac{y_{B}(x)}{x^{2}} - 2 u \phi_{i} + \phi_{i} \right ) T_{r} \; ,
\label{eq:tbr}
\end{equation}
\noindent 
and
\begin{equation}
{ \Delta T^{\rm FF}_{\rm exc}  \over T_{r} } \simeq { y_B(x) \over x^{2} } 
\label{eq:tbrexc}
\end{equation}
\noindent 
quantifies the relative excess over $T_{r}$ due to the FF diffuse emission.

In the above equations a uniform medium is assumed. This is not critical for the global Comptonization distortion because it depends linearly on matter density, but 
it introduces a significant underestimation of the FF distortion in the presence of substantial intergalactic medium (IGM) matter density contrast 
because bremsstrahlung depends quadratically on matter density. 

\begin{figure*}[!ht]
\centering
        \includegraphics[width=8.cm,angle=0]{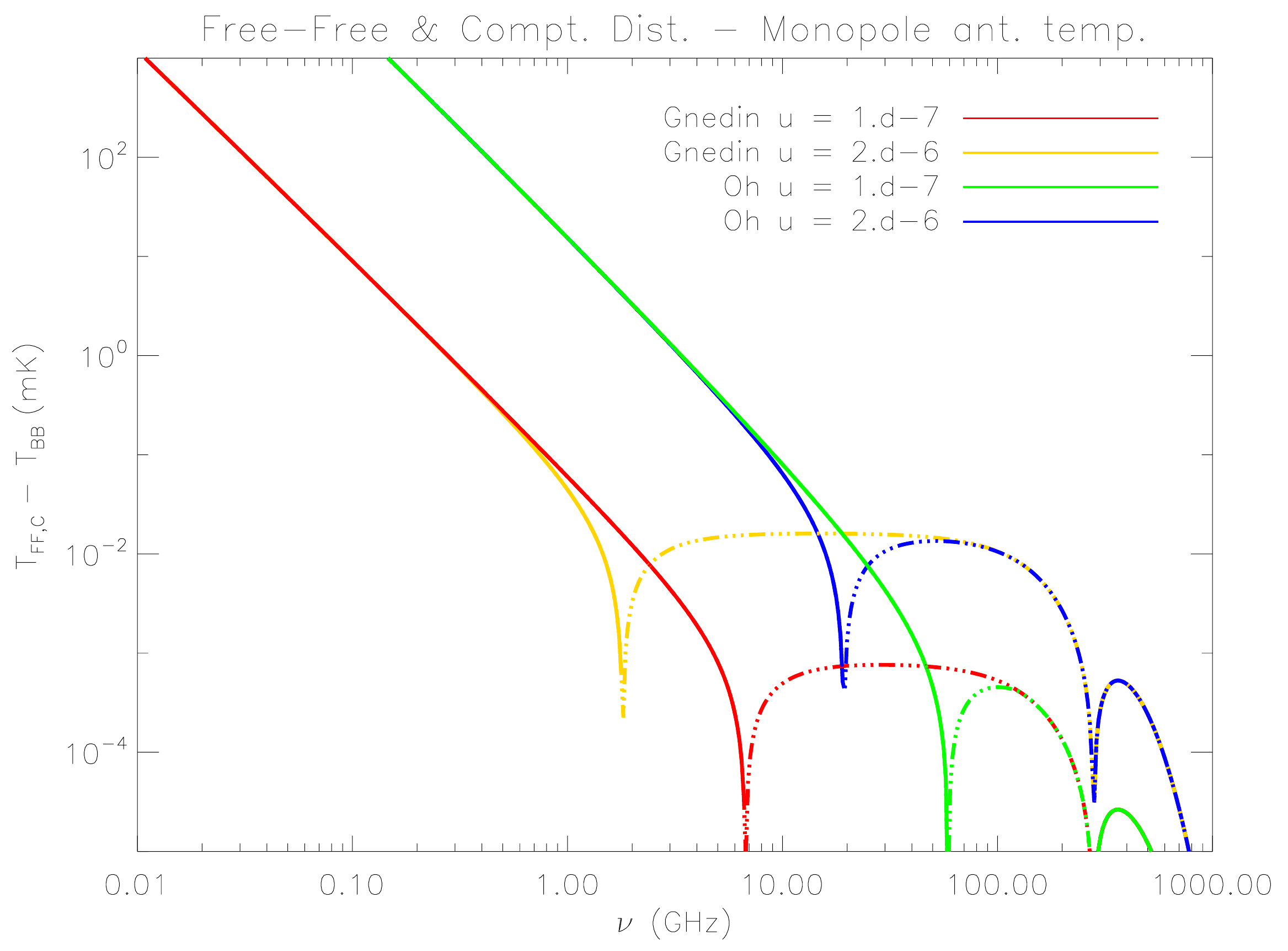}
        \includegraphics[width=8.cm,angle=0]{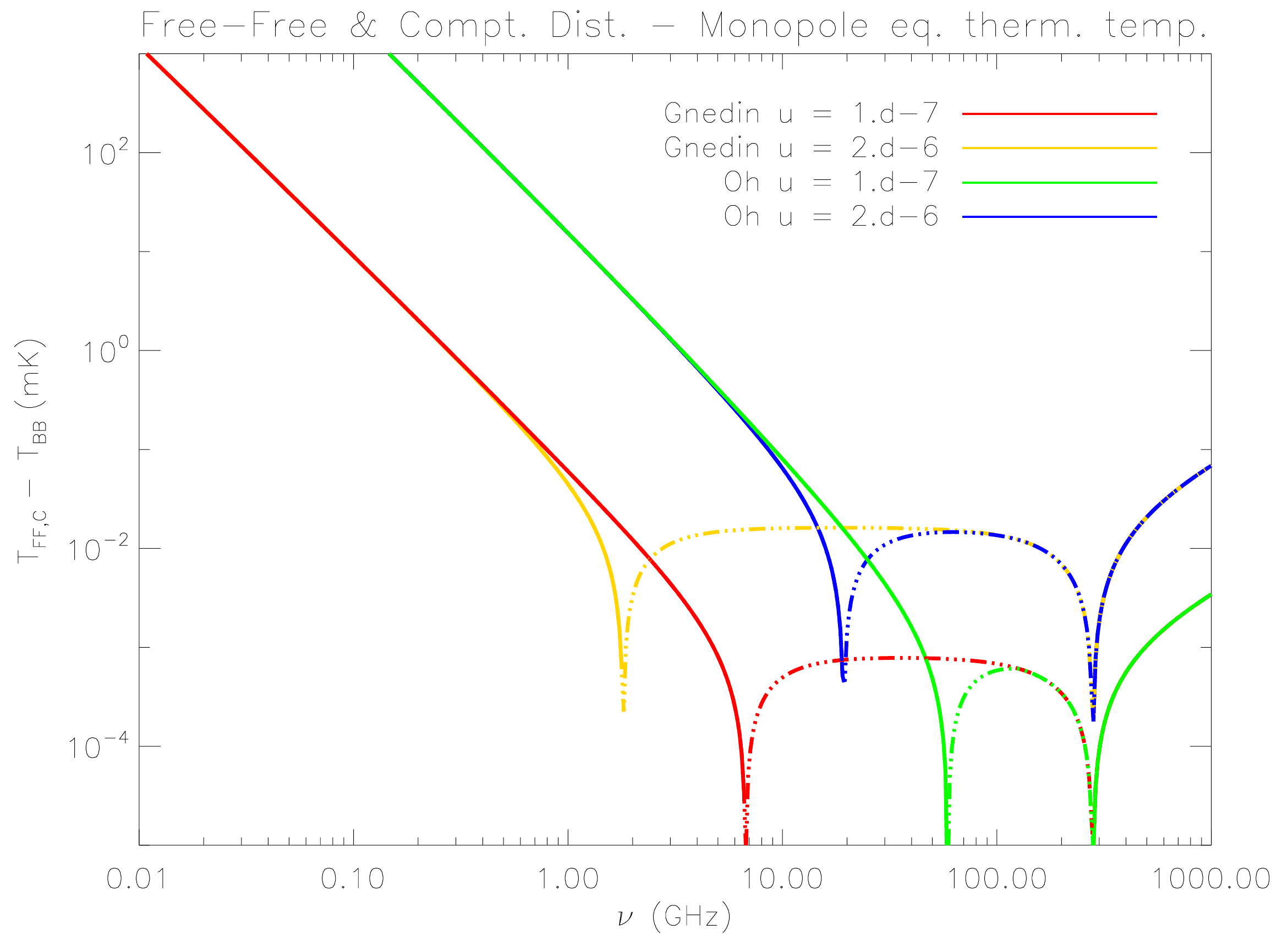}
    \caption{Monopole signal for the considered combined Comptonization and diffuse FF distortion models after subtracting the current CMB spectrum in the blackbody approximation at the effective temperature 
    $T_{0}$. Solid and three dots-dashed lines refer to positive and negative values, respectively. Free-free distortions are evaluated for a reionization model compatible with {\it Planck} results and for an extreme model involving 
    a population of ionized halos. 
    Comptonization distortions are evaluated for $u=2\times10^{-6}$
   and $u=10^{-7}$, representative of imprints
    by astrophysical or minimal reionization models.     
    For comparison with previous works that focussed on the CMB, they are also displayed, other than in terms of antenna temperature (left panel), which is typically used at radio frequencies, in terms of equivalent thermodynamic temperature
    (right panel); the differences appear above $\simeq 40$ GHz. See also the legend and the text.}
    \label{fig:MonoModFF}
\end{figure*}

A separation approach, resulting in agreement with numerical simulations \citep{2011MNRAS.410.2353P},
has been adopted to take into account inhomogeneities in matter density \citep{2014MNRAS.437.2507T}. In principle, we could also include higher order corrections from inhomogeneities in the medium temperature and differences in the 
inhomogeneities of free electrons and atoms in different ionization states. These terms are expected to be significantly smaller than the leading correction
considered here, however. This approach allows a versatile modelling of thermal and ionization histories for different underlying cosmological models.
Defining the density as the sum of a mean term and a small perturbation
and averaging over a volume representative of the Universe, a time-dependent clumping factor, 
$(1 + \sigma^{2})$,
accounts for inhomogeneities in the bremsstrahlung rate
by replacing
$\Omega_{b}^{2} (z) = \Omega_{b,homog}^{2}  (1+\sigma^{2}(z))$,
where $\Omega_{b,homog}^{2}$ defines the homogeneous case.
The baryonic matter variance, $\sigma^{2} (z)$, 
is influenced by the nature of the dark matter (DM) particles, especially at high wavenumbers ${\tilde k}$, 
while the large-scale structure distribution is almost independent of the model \citep{2007Sci...317.1527G}.
The intrinsic thermal velocities of warm DM particles 
imply a suppression of small-scale structure formation, which is efficient below their free-streaming scale, with a delay of the growth of structures \citep{2005PhRvD..71f3534V}. 
This loss of power in warm DM models can be approximated by a cold DM model with a suitable cut-off, $k_{max}$, in the typical range $\simeq (20 \div 10^3)$,
according to the thermal properties of the particles.
Thus, $\sigma^{2} (z)$ mainly depends on two crucial parameters: the amplitude and the cut-off of perturbations.

As examples, we considered two pairs of different FF and Componization distortion models.
We first adopted the FF model described in \cite{2014MNRAS.437.2507T} for the ionization history of 
\cite{2000ApJ...542..535G}, resulting in a Thomson optical depth $\tau$ that is well consistent with 
recent {\it Planck} results, and a fixed cut-off value $k_{max} = 100$. We coupled it with two different levels of Comptonization distortion, characterized by $u = 10^{-7}$, which is very close to that derived in 
\cite{2008MNRAS.385..404B} for the \cite{2000ApJ...542..535G} model and corresponds to an almost minimal energy injection that is consistent with the current constraints on $\tau$, 
and by $u = 2 \times 10^{-6}$, a value that accounts for possible additional energy injections by a broad set of astrophysical phenomena. 
While larger FF distortions (up to a factor of about 1.5--2) derive from  higher values of $k_{max}$ or from ionization histories with slightly higher values of $\tau$,  
the FF diffuse emission from an ensemble of ionized halos at substantial redshifts is expected to generate much stronger signals.  
Remarkably, the model by \cite{1999ApJ...527...16O} predicts a very strong global  (i.e. integrated over the ensemble of halos) FF signal, corresponding to 
$y_{B} \sim 1.5 \times 10^{-6}$ at $\nu \sim 2$ GHz. 
We rescaled the above FF model to $y_{B}({\rm 2\,GHz}) = 1.5 \times 10^{-6}$, coupled with Comptonization distortions with 
$u = 10^{-7}$ or $2 \times 10^{-6}$. We note that this level of global FF diffuse emission was found by \cite{1999ApJ...527...16O} by integrating the number counts of FF emitters 
over the whole astrophysical range, i.e. without any subtraction of signals by individual ionized halos. Applying a subtraction of FF emitters down to faint detection thresholds, 
as is feasible with high-resolution and deep surveys such as the Square Kilometre Array (SKA), the residual FF diffuse emission will be significantly lower.
These four models likely identify a plausible range of possible distortions. The studied range for Comptonization and FF distortions lies well within 
the current observational constraints on $u$, $|u| \le 1.5 \times 10^{-5}$ (at 95\% CL), which mainly come from the
Far Infrared Absolute Spectrophotometer (FIRAS) \citep{1996ApJ...473..576F} on board the Cosmic Background Explorer (COBE), and on $y_B$,
which is derived including previous low-frequency measurements as well \citep{2002MNRAS.336..592S} and the TRIS data \citep{2008ApJ...688...24G},
which set $-6.3\times10^{-6} <  y_B < 12.6 \times10^{-6}$ (at 95\% CL), and, obviously,
within the relaxation of $y_B$ limits, $|y_B| < 10^{-4}$ \citep{2011ApJ...734....6S},
which are mainly due to the excess \citep{2011ApJ...730..138S}  at $\simeq 3.3$ GHz in the second generation of the Absolute Radiometer for Cosmology, Astrophysics, and Diffuse Emission (ARCADE 2) data
(if not explained by a  population of faint radio sources; see
also Sect. \ref{sec:exback}).
The models are shown in Fig. \ref{fig:MonoModFF} after the current CMB spectrum in the blackbody approximation at its effective temperature is subtracted, both in terms of equivalent thermodynamic temperature and antenna temperature,
\begin{equation}
        T_{\rm ant} (\nu) ={h\nu\over k} \eta(\nu) \, .
    \label{eq:t_ant}
\end{equation}
We note the change from positive to negative values, which is predicted at a frequency ranging from a few GHz to several tens of GHz, according to the amplitude of FF emission and the value of $u$.
Almost independently of the - Comptonization - distortion level, a well-known change from negative to positive values occurs at $\simeq 280$ GHz. We note that for global distortions and referring to the current, i.e. after heating, directly observable CMB effective temperature $T_{0}$, this change of sign occurs at a frequency slightly higher than $\simeq 220$ GHz that holds for the thermal Sunyaev-Zel'dovich effect on a galaxy cluster.

At extremely low frequencies, Eq. \eqref{eq:etaC} should be replaced by a more accurate formula (see e.g. eq. (33) in \cite{1995A&A...303..323B}) for the bremsstrahlung process because the extremely high efficiency of the process implies the achievement of a blackbody spectrum at electron temperature.
On the other hand, the simple representation in terms of $y_{B}$ expressed by Eqs. \eqref{eq:etaC} and \eqref{eq:tbr}, which predicts a continuous equivalent thermodynamic temperature increase at decreasing frequency, well approximates the exact solution in the observable frequency range (see Appendix \ref{app:ybvalidity}).

At low frequencies, $y_{B}$ can be approximated by a power law,
\begin{equation}
y_{B} (\nu)  \simeq A_{\rm FF} \, (\nu/{\rm GHz})^{-\zeta} \, 
\label{eq:yBpowlaw}
,\end{equation}
\noindent 
with $\zeta \simeq 0.15$, and $A_{\rm FF} \simeq 7.012 \times 10^{-9}$ or $1.664\times 10^{-6}$ for the two considered models.

\subsection{21cm redshifted line}
\label{sec:21cm}

A precise measure and mapping of the redshifted 21cm line signal with current and next-generation radio telescopes
is likely the best way to investigate the evolution and distribution of neutral hydrogen ($HI$) in the Universe from the dark ages to the 
reionization history (epoch of reionization, EoR).

This 21cm line corresponds to the spin-flip transition in the ground state of neutral hydrogen. In cosmology,
the 21cm signal, usually expressed in terms of antenna temperature, is described as the offset of the 21cm brightness temperature from the CMB temperature, 
$T_{\rm CMB}$, along the observed line of sight at frequency $\nu$ and is given by \citep{2006PhR...433..181F}
\begin{align} 
T_{\rm ant}^{\rm 21cm} (\nu) & = \frac{T_{S} - T_{\rm CMB}}{1+z} \left (1-e^{-\tau_{\nu_{0}}} \right ) \nonumber
\\ & \simeq 27 x_{HI} \left (1-\frac{T_{\rm CMB}}{T_{S}} \right ) \left (1+\delta_{nl} \right ) \left ( \frac{H(z)}{d\nu_{r} / dr + H(z)} \right ) \nonumber \\ &
\sqrt{\frac{1+z}{10}\frac{0.15}{\Omega_{m}h_{100}^{2}}} \left (\frac{\Omega_{b}h_{100}^{2}}{0.023} \right ) {\rm mk} \, ,
\label{eqn:21cm}
\end{align}
\noindent where $T_{S}$ is the gas spin temperature, which represents the excitation temperature of the 21cm transition, $x_{HI}$ is the neutral hydrogen fraction, $\tau_{\nu_{0}}$ is the optical depth at the 21cm frequency
$\nu_{0}$, $H(z)$ is the Hubble function,
$\Omega_{m}$ is the (non-relativistic) matter density parameter, $h_{100} = H_0 / (100\, \rm{km\, s^{-1} Mpc^{-1}})$,
$\delta_{nl}(\vec{x}, z) \equiv \rho/\bar{\rho} - 1$ is the evolved (Eulerian) density contrast, 
$d \nu_{r}/dr$ is the comoving gradient of the line-of-sight component of the comoving velocity, and all the terms are computed at redshift $z = \nu_{0} / \nu -1$. The CMB is assumed as a back light, thus if $T_{S} < T_{\rm CMB}$, the gas is seen in absorption, while if $T_{S} > T_{\rm CMB}$, it appears in emission. More in general, $T_{\rm CMB}$ should include not only the CMB (blackbody) background defined by $T_r$, but also potential distortions and other possible radiation backgrounds that are relevant at radio frequencies.
Eq. \eqref{eqn:21cm} shows a dependence on the IGM temperature and ionization evolution, as well as on fundamental cosmological parameters.

Because the 21cm signal is a line transition, its frequency redshifts from each specific cosmic time during the EoR up to the current time into a well-defined frequency.
The signal detected at a given frequency then refers to a particular redshift, implying that the 21cm analysis constitutes a unique tomographic observation of the cosmic evolution. 

Different models predict different gas ionization fraction and spin temperature evolutions and, correspondingly, different predictions for the global 21cm background signal, $T_{\rm ant}^{\rm 21cm} (\nu)$, 
related to the underlying astrophysical emission sources and feedback mechanisms. 
A wide set of models has been considered in \cite{2017MNRAS.472.1915C}, resulting in an envelope of possible predictions for $T_{\rm ant}^{\rm 21cm} (\nu)$.

Recent limits on the global 21cm background signals have been set by \cite{2016MNRAS.461.2847B}, who applied a fully Bayesian method to a 19-minute-long observation from the Large aperture Experiment to detect
the Dark Ages (LEDA) to identify the faint signal from the much brighter foreground emission. 
They found a signal amplitude between $-890$ and $0$ mK (at 95\% CL)  
in the frequency range $100 > \nu > 50$ MHz, corresponding to a redshift range $13.2 < z < 27.4$. These data set limits on the 21cm signal from the cosmic dawn before or at the beginning of the increase in
electron ionization fraction and constrains structures and IGM thermal history in connection with heating sources.

Surprisingly, a pronounced absorption profile with an almost symmetric U-shape centred at $(78 \pm 1)$ MHz has recently been found by \cite{EDGESobs2018Nature}, who analyzed a set of observational campaigns that started in August 2015 and were carried out with low-band instruments of the Experiment to Detect the Global Epoch of Reionization Signature (EDGES).
The amplitude of the absorption feature was found to be $0.5^{+0.5}_{-0.2}$ K, which is more than a factor of two greater than those predicted by the most extreme astrophysical models \citep{2017MNRAS.472.1915C} that cannot account for such a deep signature. The spread of the profile has a full width at half-maximum of $19^{+4}_{-2}$ MHz. The 
low-frequency edge supports the existence of an ionizing background by 180 million years after the Big Bang, and the high-frequency edge indicates that the gas was heated to above the radiation 
temperature less than 100 million years later \citep{EDGESobs2018Nature}. 

\begin{figure} 
\centering
        \includegraphics[width=8.cm,angle=0]{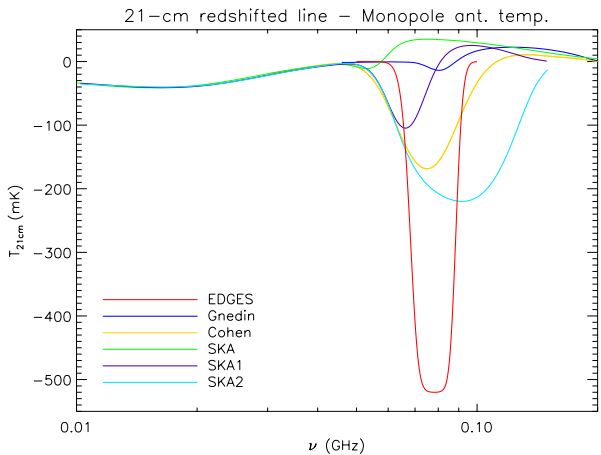}
    \caption{Monopole signal in terms of antenna temperature for the models of redshifted 21cm line.
    The depth of the EDGES profile is greater by a factor of $\simeq 3$ than that of the standard case model in \cite{2017MNRAS.472.1915C}. See also the legend and the text.}
    \label{fig:MonoMod21cm}
\end{figure}

We considered six models (displayed in Fig. \ref{fig:MonoMod21cm}): the first is the analytical representation of the EDGES absorption profile, and the other five fall in the region of the $\nu-T_{\rm ant}^{\rm 21cm}$ plane that was identified in \cite{2017MNRAS.472.1915C}. One model was presented in \cite{2008MNRAS.384.1525S} for the ionization history of 
\cite{2000ApJ...542..535G}, which we also considered for the FF diffuse emission in Sect. \ref{sec:FF}, another is the standard case model in \cite{2017MNRAS.472.1915C}. The other three models, which were also taken into account in the
study for the Square Kilometre Array (SKA) by \cite{2015aska.confE..14S} and are labelled here SKA, SKA1, and SKA2, come  
from \cite{2010PhRvD..82b3006P}, 
who exploited the dependence of 21cm signal on the X-ray emissivity 
($f_X = 0.01$ and  $f_X = 1$),
and from \cite{2012AdSpR..49..433B} (the case of a model without heating).
Fig. \ref{fig:MonoMod21cm} shows that the spectral shapes predicted for the global signal of the redshifted 21cm line present substantial differences from one model to another. 
In particular, different models predict signals with sign changes or vanishing values at different frequencies.
We note that, since the effect is important at low frequencies, antenna and equivalent thermodynamic temperatures are typically similar, but not for small values of $T_{\rm ant}$ (below some tens of mK) and particularly when it vanishes or tends to vanish.    
The differences between the redshifted 21cm line models are again different from the ones derived for the combined Comptonization and FF distortion, which at least to a good approximation simply rescales according to two global parameters. The six models considered here do not represent the full set of possibilities, but span a wide range of behaviours and amplitudes. They are exploited in the next sections to illustrate the expected types of diffuse dipole signal. We note also that they are predicted to be relevant in a limited range of frequencies. In this range, the amplitudes of the combined FF diffuse emission and of the global signal of the redshifted 21cm line span a similar and model-dependent range, but their spectral shapes are very different.

\subsection{Extragalactic background}
\label{sec:exback}

The signals discussed in Sects. \ref{sec:FF} and \ref{sec:21cm}, which are tightly related to the ionization and thermal history of IGM even though they are coupled with clumping and galaxy evolution, are intrinsically diffuse. 
An important extragalactic background signal is expected from the integrated contribution of discrete sources. A large fraction of it can be resolved through galactic surveys that reach increasingly deeper flux density levels.
In spite of this, a residual background that is observationally diffuse comes from the contribution of faint sources below the survey detection limits, and depends on their number counts.
This signal is in general very high compared to the fine signatures that directly come from reionization. Other than intrinsically interesting, this extragalactic background needs to be accurately known in order to
understand the reionization imprints correctly. Remarkably, these classes of signals may also be tightly related. 
As shown in Eq. \eqref{eqn:21cm}, the explanation of the 21cm redshifted line signal found by EDGES might require
a substantial cooling of the IGM gas (at a temperature well below that of the standard adiabatic prediction) or
an additional high-redshift extragalactic radio background (much larger than the CMB background for a blackbody spectrum at an equilibrium temperature 
equal to the temperature observed by FIRAS and rescaled by $1+z$) or a combination of them. The cooling might be driven by 
interactions between DM and baryons \citep{Barkana2018Nature,2018arXiv180210094M}.
A first claim of a notable extragalactic radio background (see also the recent radio data by \cite{Dowell:2018}) was proposed by \cite{2011ApJ...734....6S} to explain the 
signal excess found by ARCADE 2. The spectral shape of this background was too steep to be well accounted for by a fit in terms of FF emission.
In principle, this strong extragalactic radio background might be an intrinsic CMB feature \citep{2019arXiv190808876B}
or might be ascribed to a substantial astrophysical contribution. This might be generated by a faint population of radio sources \citep{2018ApJ...858L..17F}, 
for instance, or by a strong emission, with heavy obscuration or a large radio-loud fraction that are associated with the evolution of black holes and Population III stars at $z \gsim 17$
and with a growth declining at $z < 16$ \citep{2018ApJ...868...63E}. Another possibility are 
black hole high-mass X-ray binary microquasars \citep{2019arXiv190200511M}.
According to \cite{2011ApJ...734....6S}, the limits on FF distortion
significantly change depending on whether a substantial extragalactic source radio background is assumed.

We exploit several simple analytical representations of the extragalactic radio background. Some of them are adopted in the next sections to illustrate the expected types of dipole signal. 

Initially, we assumed the best-fit power-law model
\begin{equation}
T_{\rm ant}^{\rm Back} (\nu)  \simeq 18.4 \, {\rm K} \, (\nu/0.31 {\rm GHz})^{-2.57}
\label{eq:backseiff}
\end{equation}
\noindent
by \cite{2011ApJ...734....6S} (expressed in terms of antenna temperature). 

A careful analysis and prediction of the extragalactic source radio background, also including different source detection thresholds, has been carried out by \cite{2008ApJ...682..223G}. We considered their 
best-fit single power-law model for the extragalactic source background signal (again in terms of antenna temperature),
\begin{equation}
T_{\rm ant}^{\rm Back} (\nu)  \simeq 0.88 \, {\rm K} \, (\nu/0.61 {\rm GHz})^{-2.707} \, .
\label{eq:backgerv}
\end{equation}
\noindent
The authors provided an empirical analytical fit function of the differential number counts normalized to the Euclidean distribution of sources. With this recipe, we estimated the extragalactic radio background
proportional to
$\int_{S_{\rm min}}^{S_{\rm max}} S N'(\nu) dS$, where $N'(\nu)$ are the differential number counts. When a range for the parameters of the fit representation of $N'(\nu)$ is provided, we assumed their central values for simplicity.
Because the authors found a reasonable convergence of the integration by extending $S_{\rm min}$  to $10^{-12}$ Jy, 
we computed the extragalactic radio background assuming that $S_{\rm max}$ is a single parameter, to be set according to the considered source detection threshold.
We fitted the results that were computed in the frequency range between 0.151 GHz and 8.44 GHz that was investigated by the authors for certain values of $S_{\rm max}$. 
We find that a simple power law represents a good approximation, given the current uncertainties on number counts at faint flux densities.
As examples of numerical estimates, we assumed $S_{\rm max} = 50$ nJy, which almost corresponds to typical detection limits of the ultra deep reference continuum surveys planned for the SKA \citep{2015aska.confE..67P}, and $S_{\rm max} = 15$ nJy because number counts down to 
flux densities fainter than the detection threshold can be investigated through $P(D)$ methods (see \cite{2014MNRAS.440.2791V} for an analysis at 3 GHz; see also \cite{2016MNRAS.462.2934V}).
Fitting the results for the extragalactic radio background up to these flux densities,
we find
\begin{equation}
T_{\rm ant}^{\rm Back} (\nu)  \simeq A \, (\nu/{\rm GHz})^{-2.65} \, 
\label{eq:backgervres}
\end{equation}
\noindent
with $A \simeq 2.35$ and 1.51 mK. The authors also found an uncertainty from $\simeq 6$\% to $\simeq 30$\% from the lowest to highest frequencies in the above estimate.
When we assume a $\simeq 10$\% uncertainty at all frequencies in this characterization, the remaining residual extragalactic radio background
is found to be smaller by a factor $\simeq 10$ than the above estimate. 

\begin{figure} 
\centering
        \includegraphics[width=8.cm,angle=0]{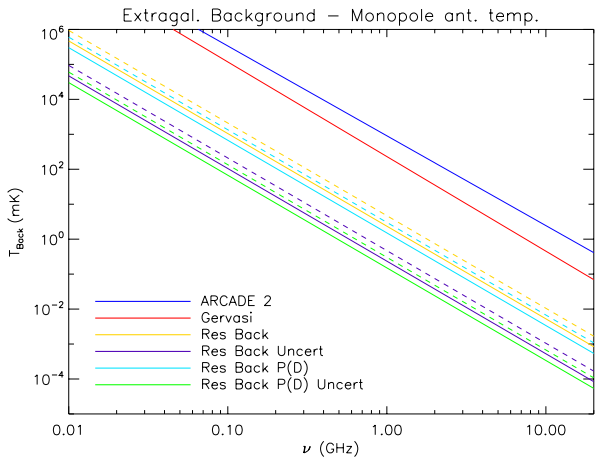}
    \caption{Monopole signal in terms of antenna temperature for the two considered extragalactic background models and various estimates of extragalactic source background signal 
    for different assumptions of source contribution subtraction. The labels "Res Back" and "Res Back P(D)" refer to the choice of $A = 2.35$ and 1.51 mK in Eq. \eqref{eq:backgervres}, while the additional label "Uncer" includes a reduction by factor of 10. The dashed lines refer to the same estimates, but derived assuming an increase of a factor of 2 in the differential number counts at faint flux densities. See also the legend and the text.}
    \label{fig:MonoModBack}
\end{figure}

The models are shown in Fig. \ref{fig:MonoModBack} (solid lines). A deeper source subtraction and a better modelling clearly imply that a lower effective extragalactic radio background 
needs to be taken into account.

Recently, the Lockman Hole Project and deep Low Frequency Array (LOFAR) imaging of the Bo\"otes field have indicated a certain flattening of differential number counts at 1.4 GHz below $\approx 100$ $\mu$Jy \citep{2018MNRAS.481.4548P} 
and at 0.15 GHz below $\approx 1$ mJy \citep{2018A&A...620A..74R}. This may suggest an increase in differential number counts of a factor of $\sim 2$ with respect to the estimate of \cite{2008ApJ...682..223G}
at the faint flux densities adopted above for computing 
the integral of $S N'(\nu) dS$,
which is likely related to the emerging of star-forming galaxies and radio-quiet AGNs at fainter flux densities.
An extragalactic radio background a factor of $\sim 2$ larger, displayed with dashed lines in Fig. \ref{fig:MonoModBack}, is directly derived in this case.

\section{Basic formalism of the differential approach}
\label{sec:diff_approach}

The peculiar motion of an observer with respect to an ideal reference frame, set at rest with respect to the CMB, produces
boosting effects in several observable quantities. The largest effect is the dipole, that is, the multipole $\ell=1$ anisotropy in the solar system
barycentre frame. 

The peculiar velocity effect on the frequency spectrum can be evaluated on the whole sky
using the complete description of the Compton-Getting effect \citep{1970PSS1825F}. This is based
on the Lorentz invariance of the photon distribution function.
At a given $\nu$, the photon distribution function, $\eta^{\rm BB/dist}$,
for the considered type of spectrum needs to be computed with the frequency multiplied by the product $(1 - \hat{n} \cdot \vec{\beta})/(1 - \vec{\beta}^2)^{1/2}$.
This accounts for all the possible sky directions, which are defined by the unit vector ${\hat{n}}$,
with respect to the peculiar velocity of the observer,
which is defined by the vector $\vec{\beta}$. This includes all the
orders in $\vec{\beta}$ and the link with the geometrical properties induced at each multipole. 
The notation `BB/dist' stands for a blackbody spectrum or for any type of non-blackbody signal, 
such as the FF plus Comptonization (C) distortion or the global signal of the 21cm redshifted line from reionization imprints or the extragalactic radio background or for combinations of them. 
In terms of equivalent thermodynamic temperature, the observed signal map is given by \citep{2018JCAP...04..021B}
\begin{equation}
T_{\rm th}^{\rm BB/dist} (\nu, {\hat{n}}, \vec{\beta}) =
\frac{xT_{0}} {{\rm{ln}}(1 / (\eta(\nu, {\hat{n}}, \vec{\beta}))^{\rm BB/dist}  + 1) } \, ;
\label{eq:eta_boost}
\end{equation}
\noindent here $\eta(\nu, {\hat{n}}, \vec{\beta}) = \eta(\nu')$ with
$\nu' = \nu (1 - {\hat{n}} \cdot \vec{\beta})/(1 - \vec{\beta}^2)^{1/2}$.

Eq. \eqref{eq:eta_boost} generalizes the result found by \cite{1981A&A....94L..33D}
for the difference in $T_{\rm th}$ measured in the direction of motion and in its perpendicular direction, expressed by
\begin{equation}\label{eq:DeltaTtherm}
    \Delta T_{\rm th}={h\nu\over k}\left\{{1\over
    \ln\left[1+1/\eta(\nu)\right]}-{1\over \ln\left[1+1/\eta(\nu(1+\beta))\right]}   \right\} \, ,
\end{equation}
\noindent
whose first-order approximation, given by 
\begin{equation}\label{eq:DeltaTtherm_firstord}
    \Delta T_{\rm th}
    \simeq -{x \, \beta \, T_0\over (1+\eta)\ln^2(1+1/\eta)}{d\ln\eta\over d\ln x} \, ,
\end{equation}
\noindent
underlines the contribution to the spectrum of the dipole that comes from the first logarithmic derivative of the photon occupation number with respect to the frequency.
Different spectral shapes of monopole signals under consideration translate into different dipole spectra.

In the next sections we compute the effect at radio frequencies 
through the Eq. \eqref{eq:DeltaTtherm} for the various monopole signals described in Sect. \ref{sec:monmod}. In principle, the difference in $T_{\rm th}$ expressed by Eq. \eqref{eq:DeltaTtherm} also includes the contributions from multipoles $\ell > 1$. 
They introduce (small) higher order corrections (see e.g. the quadrupole and octupole maps computed in \cite{2018JCAP...04..021B}), however, that we neglected in the numerical computations here.
For numerical estimates, we assumed the CMB dipole to be due to velocity effects only and set $\beta = A_{\rm dip} / T_0 = 1.2345 \times 10^{-3}$, where
$A_{\rm dip} = (3.3645 \pm 0.002)$ mK is the nominal CMB dipole amplitude according to {\it Planck} 2015 results 
\citep{2016A&A...594A...1P,2016A&A...594A...5P,2016A&A...594A...8P}. While the {\it Planck} 2018 result obtained by the Low Frequency Instrument (LFI) \citep{2018arXiv180706206P} is almost identical to the nominal {\it Planck} 2015 result, 
the High Frequency Instrument (HFI) result \citep{2018arXiv180706207P} indicates a slightly lower value, $A_{\rm dip} = 3.36208$ mK, which is still compatible within the errors.

We focus on the exploitation of the spectral shapes of the considered dipole signals. In general, see
Eq. \eqref{eq:DeltaTtherm_firstord}, the overall dipole amplitude is proportional to $\beta \, T_0$. As discussed above, this property has been used to determine (under the assumption of a CMB blackbody spectrum) the observer's motion from CMB surveys, but it can be applied to other types of backgrounds. Remarkably (see Sect. \ref{sec:monmod}), cosmic backgrounds at different frequencies 
derive from (cosmological or astrophysical) signals that are differently weighted for different redshift shells.
Because precise dipole analyses constrain our motion with respect to the background frame, the comparison between the set of our peculiar velocities with respect to different backgrounds can be used to probe the geometry and expansion of the Universe across time and to constrain the intrinsic CMB dipole. This type of analysis could represent a promising way to probe the cosmological principle, 
similarly to the investigations that are based on the comparison of the CMB dipole with dipoles from ensembles of galaxies and their number counts or distributions (see e.g. \cite{2017MNRAS.471.1045C,  2018JCAP...04..031B, 2019MNRAS.486.1350B} for studies with radio continuum surveys).

\section{Single signal results}
\label{sec:each}

\begin{figure}[ht!]
\centering
        \includegraphics[width=8.cm,angle=0]{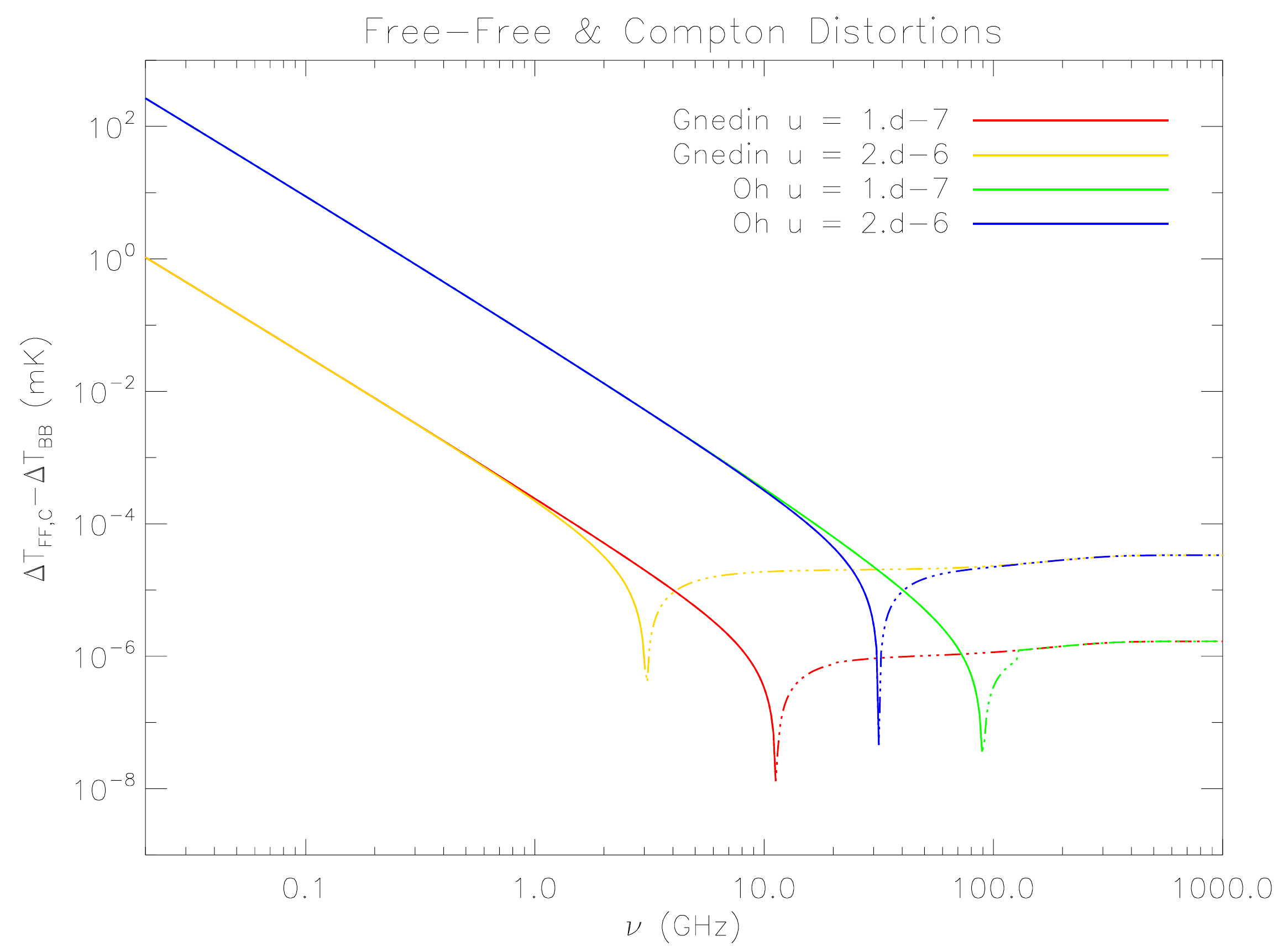}
    \caption{Dipole spectrum (in equivalent thermodynamic, or CMB,
    temperature) expressed as the difference between the dipole spectrum produced in the presence of an FF distortion (dominant at low frequencies)
    plus a Comptonization distortion (dominant at high frequencies) with monopole spectra in Fig. \ref{fig:MonoModFF} and the dipole spectrum corresponding to the blackbody at the current temperature
    $T_0$. Thick solid lines (or three dots-dashes) correspond to positive (or negative) values. See also the legend and the text.}
    \label{fig:Dip_FF}
\end{figure}

In this section we present the dipole spectra behaviours we derived for the signal. Fig.~\ref{fig:Dip_FF} illustrates the dipole spectra for 
the combination of FF and Comptonization distortion. In the plot, we subtract the "reference" dipole spectrum that corresponds to the blackbody at the current temperature $T_0$. 
The (positive) FF term dominates at low frequencies, the (negative) Comptonization term at high frequencies. The transition
from the FF to the Comptonization regime, which ranges from about 3 GHz to about 100 GHz,
depends on the relative amplitude of the two parameters $y_B$ and $u$. 
These frequencies are of potential interest both for the CMB missions that extend in frequency below the minimum foreground contamination, and for radio surveys.
We note the agreement at high frequencies with the results found in \cite{2018JCAP...04..021B} (see their Fig. 7),
where FF emission was neglected.
For $u \sim 10^{-7}$ ($u \sim 2 \times10^{-6}$) the FF term dominates at frequencies below about 10 GHz (3 GHz) for almost minimal FF models to about 100 GHz (30 GHz) for extreme FF models.

\begin{figure}[ht!]
\centering
        \includegraphics[width=8.cm,angle=0]{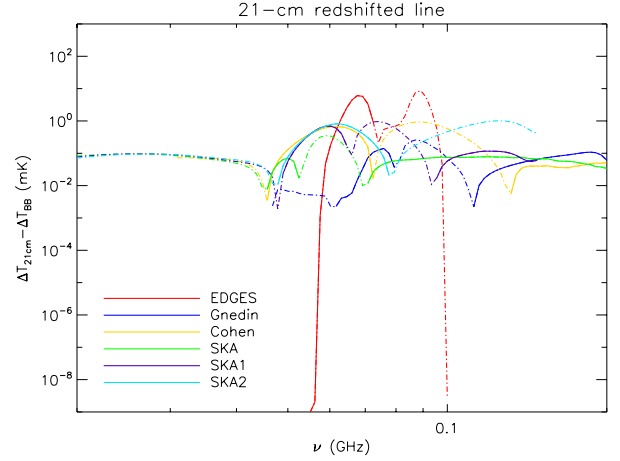}
    \caption{Dipole spectrum (in equivalent thermodynamic, or CMB,
    temperature) expressed as the difference between the dipole spectrum produced by the 21cm redshifted line signal (summed in intensity with the CMB blackbody) 
    for the monopole models in Fig. \ref{fig:MonoMod21cm}
    and the dipole spectrum corresponding to the blackbody at the current temperature
    $T_0$. Thick solid lines (or three dots-dashes) correspond to positive (or
    negative) values. See also the legend and the text.}
    \label{fig:Dip_21cm}
\end{figure}

Fig.~\ref{fig:Dip_21cm} shows the dipole spectra of the 21cm redshifted line, where we first added (in $\eta$) its 
contribution with the spectrum from the CMB blackbody at the current temperature $T_0$
(while for the case of FF plus Comptonization distortion, the contribution of the unperturbed CMB spectrum is already included in $\eta$), computed the resulting dipole, 
and then, as before, subtracted the dipole spectrum for the reference blackbody case.
For each model, we note that the changes in monopole signal behaviour are reflected in the sign inversions in the dipole spectrum (see Eq. \eqref{eq:DeltaTtherm_firstord}). 
Clearly, a larger monopole amplitude and shape steepness typically imply a larger dipole amplitude. Their combined effect is remarkable for the EDGES profile. Its dipole spectrum is 
about one order of magnitude larger than the dipole spectrum for models whose monopole signal is about two or three times smaller (compare Fig. \ref{fig:Dip_21cm} with Fig. \ref{fig:MonoMod21cm}).
The richness of the dipole profile morphology and of its sign changes might be used as diagnostic for probing the underlying model.

\begin{figure}[ht!]
\centering
         \includegraphics[width=8.cm,angle=0]{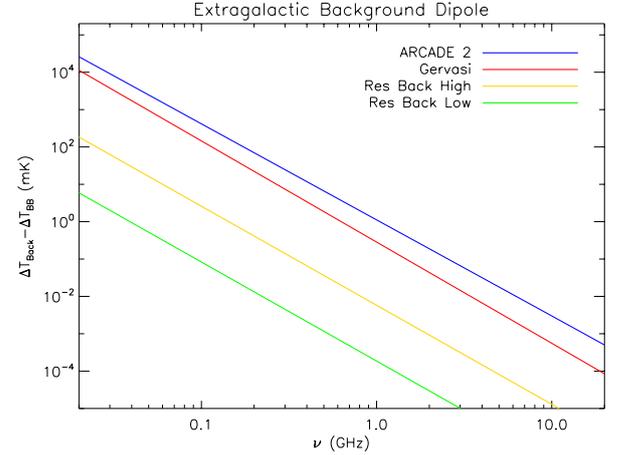}
    \caption{Dipole spectrum (in equivalent thermodynamic, or CMB,
    temperature) expressed as the difference between the dipole spectrum produced by the radio signals of the extragalactic background (summed in intensity with the CMB blackbody) 
    for the monopole models in Fig. \ref{fig:MonoModBack}
    and the dipole spectrum corresponding to the blackbody at the current temperature
    $T_0$.
    See also the legend and the text.}
    \label{fig:Dip_RadBack}
\end{figure}

In Fig.~\ref{fig:Dip_RadBack} we display the dipole spectra derived for the radio signals of the extragalactic background with the same convention adopted before.
In this representation, the simple monopole power-law dependence of the radio extragalactic background monopole is reflected in the power-law dependence of the dipole spectrum, with a very similar spectral index.
This can be simply understood by expanding Eq. \eqref{eq:DeltaTtherm_firstord} to first order in Taylor series for $\eta \gg 1$ and considering that at the frequencies of interest here, 
the CMB photon occupation number can be approximated as $1/x$. We discuss this aspect in the next section in combination with FF plus Comptonization distortions. 

We report in Fig.~\ref{fig:Dip_RadBack} the dipole spectra corresponding to the extragalactic background emission by \cite{2011ApJ...734....6S} and to the extragalactic source global radio background by \cite{2008ApJ...682..223G}. For simplicity, we only show the dipole spectra that correspond to the minimum and maximum (identified here with the labels Low and High) of the residual monopole signals in Fig. \ref{fig:MonoModBack} (i.e. from Eq. \eqref{eq:backgervres} for 
$A = 4.7$ and 0.151 mK) because they scale linearly with $A$.

\section{Signal combination results}
\label{sec:combi}

We present the dipole spectra for several combinations of signals.
The comparison between Figs. \ref{fig:Dip_FF}, \ref{fig:Dip_21cm} and \ref{fig:Dip_RadBack} clearly indicates that at radio frequencies
the dipole spectrum from the global extragalactic background (see the two top lines in Fig. \ref{fig:Dip_RadBack}) is stronger than 
the spectra from other signals.
When we considered signal combinations, we  therefore applied
source subtractions down to faint fluxes. We are mainly interested in the dipole spectrum at radio frequencies and therefore focus on the case
of Comptonization distortion with $u=2\times10^{-6}$, which is representative of the imprints that are expected in astrophysical models. For lower values of $u$, the transition from the FF diffuse emission 
to the Comptonization distortion regime translates into higher frequencies (see Fig. \ref{fig:Dip_FF}). 

\begin{figure*}[ht!]
\centering
         \includegraphics[width=18.5cm]{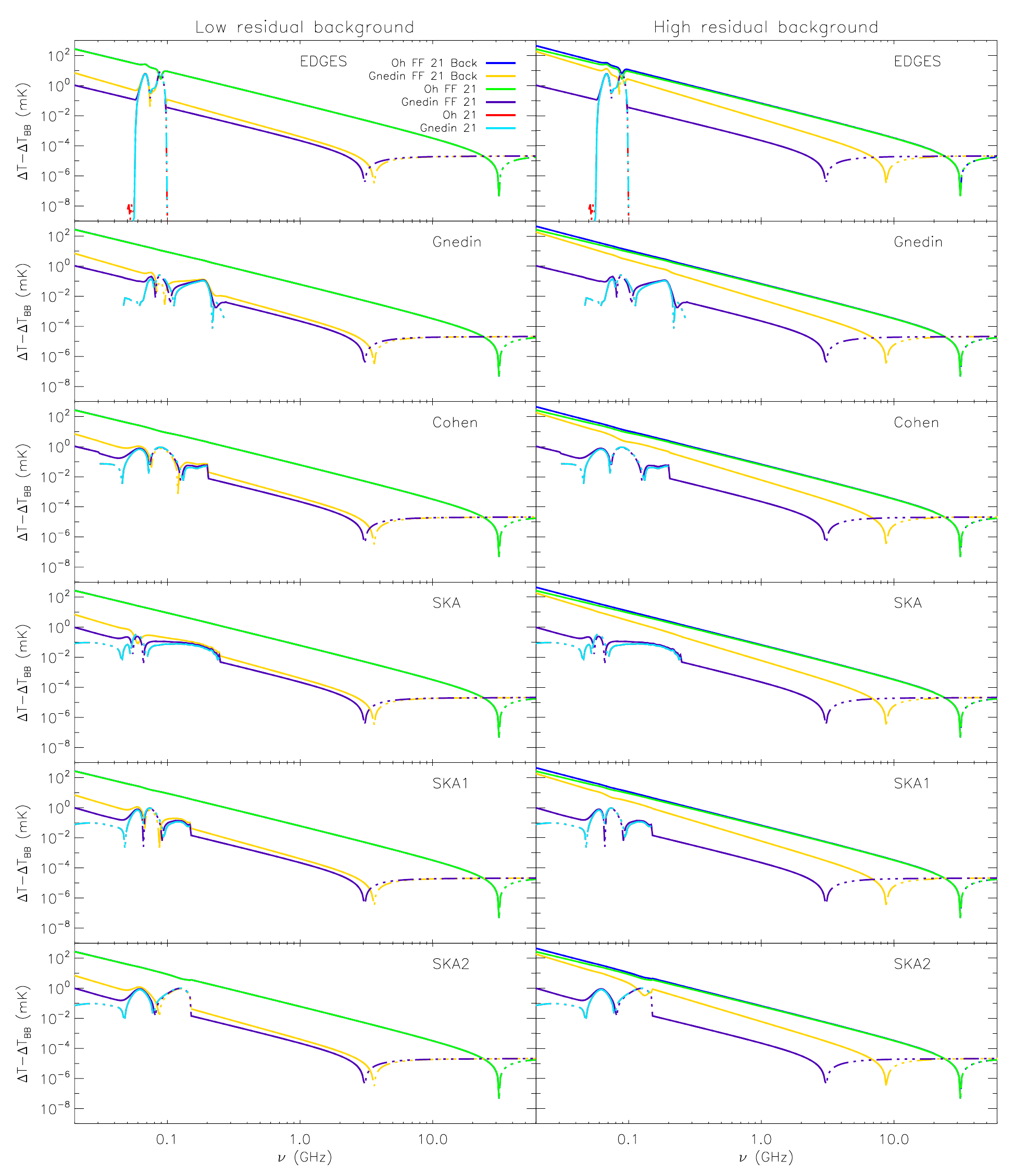}
    \caption{Dipole spectrum (in equivalent thermodynamic, or CMB,
    temperature) expressed as the difference between the dipole spectrum produced by the combinations of signals 
    and the dipole spectrum corresponding to the blackbody at the current temperature $T_0$.
    Thick solid lines (or thin three dots-dashes) correspond to positive (or negative) values.
    See also the legend and the text.}
    \label{fig:Dip_comb}
\end{figure*}

Fig. \ref{fig:Dip_comb} shows our main results. The panels in the two columns consider the 
low and high residual extragalactic background signals (see Fig. \ref{fig:Dip_RadBack}).
The label in each panel stands for the adopted 21cm redshifted line model. Different colours identify different signal combinations.
In all panels, the two discussed FF models are exploited.

The blue (yellow) lines indicate the higher (lower) FF diffuse emission plus Comptonization distortions, the 21cm redshifted line, and the residual extragalactic background,
as derived from Eq. \eqref{eq:DeltaTtherm} summing in $\eta$ the corresponding photon occupation numbers, while
in the green (violet) lines the residual extragalactic background is not included.
For a lower residual background, the dipole from the four-signal combination is almost identical to the dipole derived when the residual background was neglected
(the blue lines are masked by the green lines in the left panels).
This does not hold in the case of a higher residual background (and the yellow and violet lines are distinguishable in all the panels).

According to the model, the contribution from the 21cm redshifted line appears as a modulation of the dipole spectrum from signal combinations (see the blue and yellow lines),
which affects its approximate (see discussion below) power-law dependence at frequencies between about 60 MHz and 200 MHz. As expected, this feature is more evident for models with stronger and steeper monopole signals, and it is remarkable for the
EDGES profile. When we assume that the contribution of the residual background is subtracted, the 21cm redshifted line modulation is more or less evident over the FF emission dipole spectrum depending 
on their relative amplitudes (compare the green and violet lines). For a lower FF emission, the dipole sign inversions associated with these modulations are
appreciable as well, according to the model, in the combination of all signals and clearly better when the residual background is subtracted (see the yellow and violet lines).

When we assume that the contribution of the FF diffuse emission was also accurately subtracted, the 21cm redshifted line dipole spectrum emerges.
This is shown by the red and light blue
lines in Fig. \ref{fig:Dip_comb}, but remarkably, the main features displayed by the light blue lines are also visible in the violet lines, and even in the yellow lines for a 
lower residual extragalactic background (because they are ideally identical, the red line is masked by the light blue line; the little differences that appear at extremely low values in the case of EDGES are due only to numerical accuracy because the two curves are derived by subtraction starting from the two different global dipole predictions).
This leaves room for extracting the contribution of the dipole spectrum of the 21cm redshifted line that is embedded in the other background dipoles. This extraction is challenging, however.

In a wide radio frequency range, far from the 60--200 MHz interval, the slopes of the dipole spectra of the four-signals combination (blue and yellow lines) are between the steeper one, 
corresponding to the residual extragalactic background, and the flatter one, 
corresponding to the FF diffuse emission, according to the different amplitudes of the two types of signals.
Eq. \eqref{eq:DeltaTtherm_firstord} allows us to derive a suitable approximation for the dipole spectrum from the combination of FF diffuse emission plus Comptonization distortion and extragalactic background in the Rayleigh-Jeans region, where $\eta \gg 1$.
According to Eqs. \eqref{eq:etaCapp}, \eqref{eq:t_ant}, and \eqref{eq:yBpowlaw} and
Eqs. \eqref{eq:backseiff}-\eqref{eq:backgervres}, 
we assume
\begin{equation}\label{eq:etaAllapp} 
\eta = \frac{1-3u}{x} + b x^{-(\alpha+1)} + f x^{-(\xi+1)} \, ,
\end{equation}
\noindent
where $u$ quantifies the amplitude of Comptonization distortion, and $b$ and $\alpha$ (or $f$ and $\xi$) characterize the extragalactic background (or the FF diffuse emission); here $\alpha \simeq 2.57-2.707$ and  
$\xi = 2+\zeta \simeq 2.15$. In the limit $\eta \gg 1$, $(1+ \eta) \, {\rm ln}^2 (1+1/ \eta) \simeq 1/ \eta$. Eq. \eqref{eq:etaAllapp} implies
\begin{equation}\label{eq:devlog1} 
{d\ln\eta\over d\ln x} = - \left( { 1 + G } \right) \, ,
\end{equation}
\noindent
with
\begin{equation}\label{eq:devlog2} 
G = { b \alpha x^{-\alpha} + f \xi x^{-\xi}  \over  (1-3u) + b x^{-\alpha} + f x^{-\xi} } \, .
\end{equation}
\noindent
Finally, considering that for a blackbody spectrum at temperature $T_0$ we have simply $\Delta T_{\rm th}^{\rm BB} = \beta T_0$,
after some algebra, Eq. \eqref{eq:DeltaTtherm_firstord} gives
\begin{equation}\label{eq:DipAll}
    \Delta T_{\rm th} - \Delta T_{\rm th}^{\rm BB} \simeq F_1 + F_2 + F_3 + F_4 \, , 
\end{equation}
\noindent
where
\begin{equation}\label{eq:F1}
   F_1 =  \beta T_0 b x^{-\alpha} (1+G) \, 
\end{equation}
\begin{equation}\label{eq:F2}
   F_2 =  \beta T_0 f x^{-\xi} (1+G) \,  
\end{equation}
\begin{equation}\label{eq:F3}
   F_3 =  \beta T_0 (1-3u) G \,  
\end{equation}
\begin{equation}\label{eq:F4}
   F_4 =  -3 \beta T_0 u \, .
\end{equation}

The terms $F_1$ and $F_3$ with $u=0$ and $f=0$ give the result considering the extragalactic background alone:
\begin{equation}\label{eq:DipRBalone}
    \Delta T_{\rm th}^{\rm Back} - \Delta T_{\rm th}^{\rm BB} \simeq \beta T_0 b x^{-\alpha} (1+ \alpha) =  \beta T_{\rm ant}^{\rm Back} (1+ \alpha) \, , 
\end{equation}
\noindent
because $b x^{-\alpha} = T_{\rm ant}^{\rm Back} / T_0$.

Analogously, considering also Eq. \eqref{eq:tbrexc} (with $T_r = T_0$), the terms $F_2$ and $F_3$ with $u=0$ and $b=0$ give the result when the FF diffuse emission alone is considered:
\begin{equation}\label{eq:DipFFalone}
    \Delta T_{\rm th}^{\rm FF} - \Delta T_{\rm th}^{\rm BB} \simeq \beta T_0 f x^{-\xi} (1+ \xi) \simeq  \beta \Delta T^{\rm FF}_{\rm exc} (1+ \xi) \, , 
\end{equation}
\noindent
because $f x^{-\xi} = y_B / x^2$.

The term $F_4$ provides the result for the Comptonization distortion alone.

At low frequencies, Figs. \eqref{fig:Dip_FF} and \eqref{fig:Dip_RadBack} reflect the above relations. The same holds for  
Fig. \eqref{fig:Dip_comb}, except for the further contribution from the 21cm redshifted line that is relevant in the $\sim$ 60--200 MHz range. As discussed above, the different weights of the processes and the coupling between them,
see Eqs. \eqref{eq:devlog2}-\eqref{eq:F4}, determine the overall dipole spectrum. The condition $F_2+F_3+F_4 = 0$ with $b=0$ and the approximation $1-3u \simeq 1$ gives the solution
\begin{equation}\label{eq:Dip0FFu}
x_0^{\rm FF,C} = \left( { h \nu_0^{\rm FF,C} \over kT_0 }\right) \simeq
\left[ { {A_{\rm FF} \over 3 u} (1+\xi) \left({ h \cdot 1 {\rm GHz} \over kT_0 }\right)^\zeta }\right]^{1/\xi} \, 
\end{equation}
for the frequency of the sign change corresponding to the transition from the FF to the Comptonization regime (see Fig. \ref{fig:Dip_FF}).
Because we assumed 
$\eta \gg 1$ and used a simple power-law approximation for $y_B$, the accuracy of Eq. \eqref{eq:Dip0FFu} degrades at increasing frequency. The comparison with Fig. \ref{fig:Dip_FF},
based on Eq. \eqref{eq:DeltaTtherm} without further approximations, shows that it works well
(or overestimates $\nu_0^{\rm FF,C}$) for the lower (or for the higher) FF diffuse emission, compared to the amplitude of Comptonization distortion.

Remarkably, Fig. \ref{fig:Dip_comb} shows that for the case of a lower FF diffuse emission,  
the frequency of the sign change (at around 3 GHz in Fig. \ref{fig:Dip_FF}) when the Comptonization distortion begins to dominate
moves to higher frequencies because of the contribution from 
residual extragalactic background (compare the violet and yellow lines). This effect is clearly more evident for a higher residual extragalactic background (compare the left and right panels). An upper limit 
to this frequency of the sign change can be derived from the condition $F_1+F_3+F_4 = 0$ with $f=0$ and the approximation $1-3u \simeq 1$. This gives
\begin{equation}\label{eq:Dip0Backu}
\left( { \nu_0^{\rm Back,C} \over {\rm GHz} }\right) =  \left( { { kT_0 \over h } x_0^{\rm Back,C} }\right) \lsim 
\left[ { {A (1+\alpha) \over 3 T_0 u}  }\right]^{1/\alpha} \, .
\end{equation}
For a much higher FF diffuse emission this frequency shift becomes almost negligible. 

With a wide frequency coverage, it would be possible to appreciate the different shapes of the various dipole signals.
Microwave surveys are particularly important for assessing the contribution of the Comptonization distortion to the 
dipole spectrum, thus improving the modelling of the FF term at lower frequencies. A certain frequency shift to higher values in the transition of the Comptonization distortion regime
can also be expected for lower values of $u$ (see Fig. \ref{fig:Dip_FF}) because of the contribution from the residual background.
Towards microwave frequencies, extragalactic sources with an almost flat spectrum (that are poorly described by the adopted number counts) become increasingly relevant.
They can be directly extracted by analysing CMB maps from the next-generation space missions or using deeper multi-frequency ground-based surveys (see e.g. \cite{2018JCAP...04..020D} and references therein
for a deeper discussion).

\section{Discussion and conclusion}
\label{sec:conclu}

The best way to study the dipole spectrum of cosmic diffuse emissions should be based on the analysis of surveys that cover almost the entire sky or provide very wide coverage.
On the other hand, investigations at microwave to sub-millimetre frequencies 
take advantage from avoiding sky regions that are mostly affected by foreground signals, even though the observed area is partially reduced. 

Surveys at radio frequencies are performed through "classical" mapping
from single telescopes or with interferometric techniques involving very many telescopes. These surveys are in particular aimed at improving resolution and sensitivity at small scales.

The SKA (see \cite{SKATELSKODD001Revision1} for specifications) frequency coverage jointly offered by the low- ($\simeq 0.05-0.35$ GHz) and medium- ($\simeq 0.35-14$ GHz) 
frequency and survey-mode ($\simeq 0.65-1.67$ GHz) arrays during the SKA-1 phase is particularly suitable when compared to the frequency location of the 21cm redshifted line, radio background, and FF signals, and also, 
depending on the models, of the transition from FF or residual background to Comptonization dominance. Frequency extensions to about 20 GHz that have been studied for the SKA-2 phase and will hopefully be expanded to higher frequencies
in a third SKA phase might allow observing this transition even for models with relatively high FF or residual background (that may come from sources with flatter spectra).
The combination of SKA resolution and sensitivity will allow us to detect point sources and to probe their number counts down to very faint flux densities (see Sect. \ref{sec:exback}), which will enable substantial subtractions of their contribution to the radio background.

Depending on the implemented configuration (i.e. the adopted baselines), interferometers are intrinsically sensitive to sky signal variations up to certain spatial scales (typically, of several degrees),
thus producing a wide area map that essentially independently collects different patches, and preventing the analysis of sky geometrical properties at angular scales larger than the patch. The map might in principle be reconstructed with suitable array designs and observational approaches, such as a compact configuration implementing short baselines and a high uv-space filling factor, mosaicking techniques, and methods such as on-the-fly-mapping for the largest areas. Furthermore, it is possible to recalibrate different sets of interferometric maps over wide area surveys that were taken with classical (single-dish) approaches to recover the information at large scales but keeping interferometric sensitivity and resolution at small scales. 

If these approaches that are useful for a broad set of SKA scientific projects work successfully, the analysis of diffuse cosmic dipoles at radio frequencies could be pursued with methods analogous to those suitable for recent CMB surveys and future missions \citep{2018JCAP...04..021B}.
The analysis of these aspects and of the applications of those methods, readapted to the radio domain, as well as the investigation of specific techniques to deal with 
independent patches from interferometric observations will be the object of a future study. 

Conversely, the dipole pattern amplitude might be reconstructed even without coherent mapping
up to the largest angular scales. Along a meridian in a reference frame where the $z$-axis is parallel to the dipole direction, 
the signal variation at colatitude $\theta$ from a dipole pattern with $\Delta T,$ see Eq. \eqref{eq:DeltaTtherm}, within a limited sky area of linear size $\Delta \theta$,
has an amplitude $|\Delta T_{\Delta \theta}| \simeq \Delta T \cdot (\Delta \theta / 90^\circ) \, {\rm sin} \, \theta$, with ${\rm sin} \, \theta$ not far from unit
 at angles large enough from the poles.

For example, at $\nu \sim (50 - 100)$ MHz, we find for the FF and 21cm signatures that the relevant signals (after the standard CMB blackbody spectrum dipole was subtracted) have amplitudes 
$ \Delta T_{\rm th} - \Delta T_{\rm th}^{\rm BB} \sim 0.1 - 50$ mK
depending on the model. Similar values also hold for the radio extragalactic background when the deepest detection thresholds are considered.
Values of about 2 (or 4) orders of magnitude higher are achieved for shallower detection limits (or for the best-fit radio signal of the extragalactic background). Thus, for a patch of size $\Delta \theta \sim 3^\circ$,
sensitivities to the diffuse signal of $\text{about}$ several tens of mK (or in a range from $\text{about}$ a few $\mu$K to $\text{about}$ mK) could allow identifying 
the extragalactic background (or the reionization imprints on the diffuse radio dipole). All these sensitivity levels can certainly be achieved based on the SKA specifications.
Subtracting sources by applying a broad set of detection thresholds
would also help to clarify to what extent the radio extragalactic background could be ascribed to extragalactic sources or if it is of intrinsic cosmological or diffuse origin. This would contribute to answering the question about the level and origin of the radio extragalactic background, which is still controversially discussed (see e.g. \cite{Subrahmanyan:2013,2018Natur.564E..32H,2018MNRAS.481L...6S}). Clearly, collecting many patches would improve the statistical information. 

The comparison of Figs. \ref{fig:MonoModFF}, \ref{fig:MonoMod21cm}, and \ref{fig:MonoModBack} with Figs. \ref{fig:Dip_FF}, \ref{fig:Dip_21cm}, and \ref{fig:Dip_RadBack} shows that the dipole amplitude, $\Delta T_{\rm th}$, that is predicted for each specific signal is significantly smaller (by two or three orders of magnitude) than the corresponding global monopole signal. This shows that its precise absolute measurement does not require 
significant corrections for dipole effect.
In contrast, the different amplitude of the dipole from different types of signal also requires a precise correction for the dipole pattern
from stronger signals, in order to achieve a precise absolute determination of the monopole of weaker components. Remarkably, this is the case of the 21cm redshifted line (compare Fig. \ref{fig:MonoMod21cm} with Figs. \ref{fig:Dip_FF} and \ref{fig:Dip_RadBack} at $\nu \sim (50 - 150)$ MHz). The level of correction depends on the intrinsic amplitude of the stronger (FF or radio background) component and on the capability of subtracting to
faint detection thresholds the (ionized halos or extragalactic) sources that are responsible for a large portion of this dominant contribution. Again, a good frequency coverage would allow us to distinguish the different signals by exploiting their different spectral shapes.

In general, an ultra-accurate subtraction of the Galactic foreground emission is the most critical issue for analysing fine cosmological signals. Differently from the Galactic foreground, 
all the signal variations in the patches from cosmic diffuse dipoles should be described by a well-defined pattern related to the motion of the observer. This would improve a joint analysis.

The study of the dipole spectrum, which links monopole and anisotropy analyses, relies on the quality of interfrequency and relative data calibration. This by-passes the need for precise absolute calibration, which is a delicate task in cosmological surveys and in particular, 
in the combined analysis of data from different experiments and of maps from radio interferometric facilities. It might therefore also represent a way to complement and cross-check investigations on cosmological reionization, 
based on the study of the global signal and of the fluctuations projected onto the sphere, or in redshift shells performed with angular power spectrum or power spectrum analyses
and higher order estimators.

\begin{acknowledgements}
It is a pleasure to thank Gianni Bernardi, Gianfranco de Zotti, and Isabella Prandoni for useful discussions.  
We also thank the anonymous referee for comments that helped improve the paper.
We gratefully acknowledge financial support from the INAF PRIN SKA/CTA project FORmation and Evolution of Cosmic STructures (FORECaST) with Future Radio Surveys, from ASI/INAF agreement n.~2014-024-R.1 for the {\it Planck} LFI Activity of Phase E2 and from the ASI/Physics Department of the University of Roma--Tor Vergata agreement n. 2016-24-H.0 for study activities of the Italian cosmology community.
TT acknowledges financial support from the research program RITMARE SP3 WP3 AZ3 U02 and the research contract SMO at CNR/ISMAR during the finalization of this work.
\end{acknowledgements}

%

\appendix

\section{Estimate of $\nu_{B}$}
\label{app:ybvalidity}

We discuss here the frequency range validity of the FF distortion approximation in terms of the parameter $y_B$ (see Eqs. \eqref{eq:free},  \eqref{eq:etaC}, and \eqref{eq:tbr}).
It holds at dimensionless frequencies
$x_B \ll x \ll 1$ (see \cite{1995A&A...303..323B} and references therein), where $x_B$
is such that $y_{abs,B} (x_B)= 1$, being
\begin{equation}
y_{abs,B} (x, \phi(z), z) = \int_1^{1+z} {t_{exp} d(1+z') \over t_{abs,B} (1+z')} \, ,
\end{equation} 
where $t_{abs,B}$ is the bremsstrahlung absorption time 
\begin{equation}
{1 \over t_{abs,B}} = K_B (z) {g(x,\phi) \over (x/\phi)^3} {\rm e}^{-x/\phi} ({\rm e}^{x/\phi} -1) \, .
\label{eq:tabs}
\end{equation} 
\noindent
The expansion time, including the contributions of the different types of energy densities in a Friedmann-Lema\^itre-Robertson-Walker model, can be expressed by 

\begin{align}
& t_{exp} = {a \over \dot{a}} = {1 \over H(z)} \nonumber \\
& = {1 \over
\left[H_0 \Omega_{rel}^{1/2} (1+z)^2\right] 
\left[ 1 + {\Omega_m / \Omega_{rel} \over 1+z} \, \left( 1 + {\Omega_K / \Omega_m \over 1+z} + {\Omega_\Lambda / \Omega_m \over (1+z)^3}  \right)  \right]^{1/2} 
} \, ,
\end{align} 

\noindent
where $\Omega_{rel}$, $\Omega_{K}$ , and $\Omega_{\Lambda}$ are  
the relativistic particle (typically radiation plus relativistic neutrinos) density parameter, 
the curvature density parameter, and the cosmological constant -- or dark energy, or vacuum energy -- density parameter.

At $x \ll 1$, in Eq. \eqref{eq:tabs}
$[{g(x,\phi) / (x/\phi)^3}] {\rm e}^{-x/\phi} ({\rm e}^{x/\phi} -1) \simeq {g(x,\phi) \phi^2 / x^2} $. Thus, the condition $y_{abs,B} (x_B)= 1$ becomes
\begin{align}
& x_B^2 \simeq  \int_1^{1+z} {t_{exp} \over (1+z')} K_B (z') g(x_B,\phi) \phi^2 \, d(1+z') \nonumber \\
& \simeq 
1.6 \cdot 10^{-7} (T_0 / 2.7 {\rm K})^{-7/2}  \nonumber \\
& \cdot  \int_1^{1+z} 
{ {{\Omega}_{b}}^2 \, h_{50}^3 \, \chi_e^2 \, \phi^{-3/2} \, g(x_B,\phi) \, d(1+z') 
\over
\Bigl\{
{ \Omega_{rel} (1+z')
\left[ 1 + {\Omega_m / \Omega_{rel} \over 1+z'} \, \left( 1 + {\Omega_K / \Omega_m \over 1+z'} + {\Omega_\Lambda / \Omega_m \over (1+z')^3}  \right)  \right] }\Bigr\}^{1/2} 
}
\, .
\end{align}
\noindent
For simplicity, we first consider in the integral only the terms in the denominator with explicit dependence on $z'$; we later discuss the effect of the other evolving terms.
In a reionization context, $z' < $ few tens and then $\Omega_{rel} (1+z') \ll 1$. When we neglect the curvature term, which is significantly constrained by recent data and affects the result by less than a few percent, 
the integral in the above equation therefore becomes
$\sim \Omega_m^{-1/2} \int_1^{1+z} (1+z')^{3/2} d(1+z') / [(1+z')^3 + \Omega_\Lambda / \Omega_m]^{1/2} \lsim \Omega_m^{-1/2}  \, z$. An upper limit to $x_B$ is then
\begin{align}
x_B \lsim 4 \cdot 10^{-4} (T_0 / 2.7 {\rm K})^{-7/4} {{\Omega}_{b}} \, \Omega_m^{-1/4} \, h_{50}^{3/2} \, \chi_e \, \phi^{-3/4} \, g(x_B,\phi)^{1/2} \, z^{1/2} \, .
\label{eq:xB}
\end{align}
For cosmological parameters that are compatible with current data \citep{2018arXiv180706209P},
Eq. \eqref{eq:xB} implies $x_B \lsim {4 \cdot 10^{-5}} \chi_e \, \phi^{-3/4} \, g(x_B,\phi)^{1/2} \, z^{1/2}$, which in terms of observational frequency, corresponds 
to $\nu_B / [\chi_e \, \phi^{-3/4} \, g(x_B,\phi)^{1/2}] \lsim 10 \, {\rm MHz}$ (or $\lsim 3.2$ MHz) for $z \simeq 20$ (or 2).
Because $g(x_B,\phi) \approx 1$ and $\chi_e \lsim 1$, possible relevant modifications to this estimate can come only from the evolution of $\phi$. 
The adopted value of $\Omega_{b}$ corresponds to a homogeneous medium, 
and this should be multiplied by $(1+\sigma^{2}(z))^{1/2}$ because of inhomogeneities (see the discussion in Sect. \ref{sec:FF}).
This factor could significantly increase $\nu_B$ only at $z \lsim $ a few units and up to at most one order of magnitude. This is largely 
compensated for by the former effect due to the electron temperature evolution (the term $\phi^{-3/4}$), which for typical reionization thermal histories with $T_e \sim 10^4$ K 
implies a decrease in $\nu_B$ by more than two orders of magnitude at $z \lsim 6$. At observational frequencies $\gsim 10$ MHz that are relevant for radio observations, 
the approximation of the FF distortion in terms of $y_B$ is therefore reliable for most models. 

\end{document}